\documentclass[draft,prl,twocolumn,showpacs]{revtex4}

\usepackage{amsmath,amsfonts,amssymb,bm}
\usepackage{dcolumn}
\usepackage[final]{graphicx}
\usepackage{bm}
\usepackage{comment}

\includecomment{pdffigure}

\begin{document}

\bibliographystyle{apsrev}

\title{Phonon-assisted decoherence in the production of polarization-entangled photons in a single semiconductor quantum dot}

\author{Ulrich Hohenester}\email{ulrich.hohenester@uni-graz.at}
\author{Gernot Pfanner}
\author{Marek Seliger}
\affiliation{Institut f\"ur Physik,
  Karl--Franzens--Universit\"at Graz, Universit\"atsplatz 5,
  8010 Graz, Austria}

\date{\today}

\begin{abstract}

We theoretically investigate the production of polarization-entangled photons through the biexciton cascade decay in a single semiconductor quantum dot. In the intermediate state the entanglement is encoded in the polarizations of the first emitted photon and the exciton, where the exciton state can be effectively ``measured'' by the solid state environment through the formation of a lattice distortion. We show that the resulting loss of entanglement becomes drastically enhanced if the phonons contributing to the lattice distortion are subject to elastic scatterings at the device boundaries, which might constitute a serious limitation for quantum-dot based entangled-photon devices.

\end{abstract}



\maketitle

 
Single quantum dots provide a viable source for entangled photons \cite{benson:00,akopian:06,stevenson:06}, as needed for quantum computation \cite{knill:01} and quantum cryptography \cite{gisin:02}: a biexciton, consisting of two electron-hole pairs with opposite spin orientations, decays radiatively through two intermediate optically active exciton states. If the exciton states are degenerate, the two decay paths of the cascade differ in polarization but are indistinguishable otherwise. Therefore, the emitted photons are entangled in polarization \cite{benson:00}. Such ideal performance is usually spoiled by the electron-hole exchange interaction \cite{gammon:96,bayer:02}, that splits the intermediate exciton states by a small amount and attaches a which-path information to the photon frequencies. This process deteriorates the photon entanglement. Spectral filtering \cite{akopian:06}, energetic alignment of the exciton states by means of external magnetic \cite{stevenson:06} or electric \cite{gerardot:07} fields, or growth optimization \cite{seguin:05} allow in principle to erase this information and to recover a high degree of entanglement. While all these schemes mask the which-path information to an outside observer, who could gain information about the photon polarization by simply measuring its frequency, it might also be the solid-state environment within which the quantum dot is embedded (e.g., lattice degrees of freedom or charging centers) that acts as an ``observer'' of the intermediate exciton state and thereby diminishes the entanglement degree.
 
It is the purpose of this Letter to theoretically investigate the loss of photon entanglement due to couplings to the solid state environment. We consider the situation that the interaction of the exciton with the lattice degrees of freedom, i.e., phonons, depends slightly on the exciton polarization. This is expected for self-organized quantum dots owing to the strong strain-induced piezoelectric fields \cite{seguin:05,bester:05} that are also responsible for the fine-structure splitting. We show that the biexciton cascade decay is accompanied by the formation of a polaron \cite{mahan:81,borri:01,krummheuer:02,hameau:99,verzelen:02}, i.e., a lattice distortion in the vicinity of the quantum dot, leading to a partial loss of coherence and photon entanglement. This effect becomes drastically magnified for single-dot devices, such as mesas or microcavities. The phonons contributing to the polaron can scatter elastically at the device boundaries \cite{rudin:06}. In consequence, the phonons continuously ``measure'' the exciton state and {\em einselect}\/ \cite{zurek:03} the system from a maximally entangled state to a mixed one. We show that this effect leads to a significant reduction of the degree of photon entanglement, and might constitute a serious limitation for quantum-dot based entangled-photon devices.

\begin{figure}[b]
\begin{pdffigure}
  \centerline{\includegraphics[width=0.9\columnwidth]{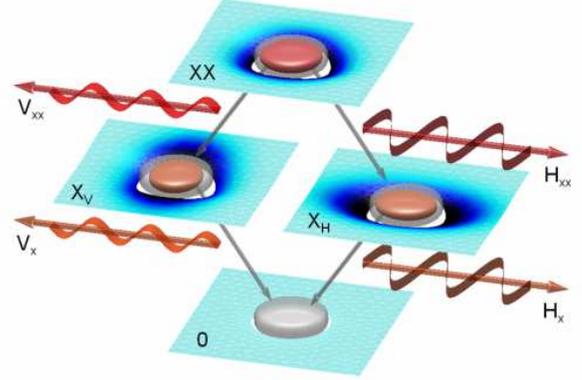}}
\end{pdffigure}
\caption{(color online) Schematic sketch of the creation of polarization entangled photons in a single semiconductor quantum dot through the cascade decay of the biexciton $X\!X$, as described in text. The two intermediate exciton states $X_H$ and $X_V$ have dipole moments oriented along $x$ and $y$, respectively; $0$ is the quantum dot groundstate. The different panels report the excitonic wavefunctions and the distortion of the lattice in the vicinity of the quantum dot, which is assumed to be slightly asymmetric for states $X_H$ and $X_V$. $H_{X\!X}$ ($H_{X}$) and $V_{X\!X}$ ($V_{X}$) denote the polarizations of the photons emitted in the biexciton (exciton) decay. We assume a well-defined polarization of the emitted photons, which can be achieved by placing the quantum dot into a microcavity \cite{benson:00,akopian:06,stevenson:06}, and equal magnitude of the dipole moments. The frequencies of the photons created in the biexciton and exciton decay differ by a few meV owing to the biexciton binding energy. }
\end{figure}


In our theoretical approach we consider a single quantum dot which is initially in the biexciton state $|3\rangle$ (two electron-hole pairs with opposite spin orientations \cite{benson:00,akopian:06,stevenson:06}, $X\!X$ in Fig.~1). In a typical experiment, the system will eventually pass after photoexcitation or electrical injection of electrons and holes through this state. It thus suffices to consider this initial condition. The biexciton is optically coupled to the two exciton states $|1\rangle$ ($X_H$) and $|2\rangle$ ($X_V$) which are polarized along $x$ and $y$, respectively \cite{bayer:02,seguin:05,bester:05}. We presume that the fine-structure splitting due to the anisotropic electron-hole exchange interaction is corrected for by appropriate external magnetic or electric fields \cite{stevenson:06,gerardot:07}, such that the energies $E_1$ and $E_2$ are identical. In the ideal case, the cascade decay
\begin{equation}\label{eq:entanglement}
  |3\rangle\to
  |H1\rangle+|V2\rangle\to
  |HH\rangle+|VV\rangle
\end{equation}
then generates a maximally entangled two-photon state \cite{benson:00,troiani:06a,troiani:06b}, where for simplicity we have suppressed all phase and normalization factors. Under realistic conditions, however, the state of the emitted radiation is affected by a number of uncertainties, such as the photon emission time (time jitter \cite{kiraz:04}) or environment couplings. In particular, if the two states $1$ and $2$ couple differently to the solid state environment, for reasons detailed below, the intermediate entangled photon-exciton state will gradually decay, with a given rate $\gamma$, into a mixed state. To quantify the resulting entanglement loss, we introduce the two-photon correlation function~\cite{mandel:95} $G^{(2)}(t,\tau)=\left<\,:\!I_{X\!X}(t) I_X(t+\tau)\!:\,\right>$, describing the polarization correlations between the photons emitted in the biexciton decay at time $t$ and the exciton decay at time $t+\tau$, respectively. It can be computed by means of the quantum regression theorem \cite{mandel:95,troiani:06a}. In the two-photon subspace spanned by the basis $HH$, $HV$, $VH$, and $VV$, we then obtain
\begin{eqnarray}\label{eq:two-photon}
  G^{(2)}(t,\tau)
  &\propto&e^{-2\Gamma t-\Gamma\tau}\,
  \left(\begin{array}{cccc}\frac 12 & 0 & 0 &\frac{e^{-\gamma\tau}}2 \\
  0 & 0 & 0 & 0\\ 0 & 0 & 0 & 0\\ \frac{e^{-\gamma\tau}}2 & 0 & 0 & \frac 12 \\
  \end{array}\right)\,,
\end{eqnarray}
with $\Gamma$ the radiative decay rate. The first exponential on the right-hand side accounts for the gradual buildup of the two-photon state due to the radiative biexciton and exciton decay, while the matrix describes the (conditional) two-photon density matrix. For the specific form given in Eq.~\eqref{eq:two-photon} one can easily check that the {\em entanglement of formation}\/ or {\em concurrence}\/ \cite{wooters:98}, which provides a quantitative measure of the photon entanglement, is simply given by $e^{-\gamma\tau}$. Thus, the longer the intermediate exciton interacts with the environment, the more the entanglement becomes diminished. The overall degree of polarization entanglement $1/(1+\gamma/\Gamma)$ is obtained through an average over all times $t$ and $\tau$ \cite{troiani:06a}.


Having established the consequences of polarization-dependent exciton couplings, we shall now propose a microscopic model for such asymmetry. Our starting point is provided by the usual {\em independent Boson model}\/ \cite{mahan:81,krummheuer:02}
\begin{equation}\label{eq:indboson}
  H=H_0+\sum_{i=1}^3\sum_\lambda\left({g_\lambda^i}^* a_\lambda^{\phantom\dagger}
  +g_\lambda^i a_\lambda^\dagger\right)|i\rangle\langle i|\,,
\end{equation}
which describes the coupling of the biexciton and the excitons to the phonons. Here, $H_0=\sum_{i=1}^3 E_i |i\rangle\langle i|+\sum_\lambda\omega_\lambda a_\lambda^\dagger a_\lambda^{\phantom\dagger}$, with $\lambda$ labeling the different acoustic phonon modes of energy $\omega_\lambda$, $a_\lambda^\dagger$ are the bosonic creation operators, and $g_\lambda^i$ the (bi)exciton-phonon couplings. Since the biexciton approximately consists of two excitons with opposite spin orientations, we set $g_\lambda^3=g_\lambda^1+g_\lambda^2$. In an ideal quantum dot the coupling constants $g_\lambda^1=g_\lambda^2$ are given by the form factors of the exciton wavefunctions \cite{krummheuer:02,hohenester.jpb:07} and the phonon coupling constants of the bulk semiconductor. These are entirely real for deformation potential coupling and entirely imaginary for piezoelectric couplings \cite{mahan:81}. For self-organized InAs/GaAs or InGaAs/GaAs quantum dots strain fields result in strong piezoelectric fields \cite{seguin:05,bester:05}, which are in part responsible for the large observed fine-structure splittings, and partially mix the deformation potential and piezoelectric couplings. Although the strength (and even sign) of these internal fields is still a matter of intense debate \cite{seguin:05,bester:05,ediger:07} and the calculation of the coupling asymmetry beyond the scope of this paper, we shall make the reasonable assumption that the phonon-exciton matrix elements differ by a small phase factor $\eta$ viz. $g_\lambda^2=e^{i\eta}\,g_\lambda^1$. 

\begin{figure}
\centerline{\includegraphics[width=0.85\columnwidth]{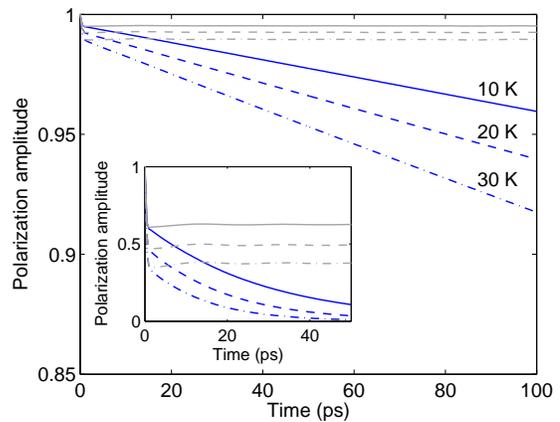}}
\caption{(color online) Decay of polarization amplitude $\left|\langle 2|\rho(t)|1\rangle\right|$ in presence of phonon couplings and for three different temperatures. At time zero the system is prepared in the superposition state $|1\rangle+|2\rangle$. The inset shows for comparison the time evolution of the optically induced polarization $\left|\langle 1|\rho(t)|0\rangle\right|$ (initial state $|0\rangle+|1\rangle$), that is directly accessible in four-wave mixing experiments \cite{borri:01}. The gray lines correspond to simulations where elastic phonon scatterings are ignored, the dark lines to simulations with $\gamma_d=50\,\mu$eV \cite{rudin:06}. In the calculations we assume an asymmetry $\eta=0.1$ and Gaussian exciton wavefunctions with a FWHM of 5 nm and 1.5 nm along the lateral and growth directions, respectively \cite{hohenester.jpb:07}.}
\end{figure}

We shall show next that even small asymmetries $\eta$ can result in a significant entanglement loss. In doing so, we exploit the fact that the independent Boson model can be solved analytically \cite{mahan:81}. Since no transitions between different excitonic states are induced by the phonon coupling \eqref{eq:indboson}, the time evolution operator is diagonal \cite{viola:98,hohenester.jpb:07}
\begin{equation}\label{eq:evolve}
  U(t)=e^{-iH_0t}\sum_i \prod_\lambda e^{i|\alpha_\lambda^i|^2\sin\omega_\lambda t}
  D_\lambda(\alpha_\lambda^i[1-e^{i\omega_\lambda t}])|i\rangle\langle i|\,,
\end{equation}
with $\alpha_\lambda^i=g_\lambda^i/\omega_\lambda$ and $D_\lambda$ the usual bosonic displacement operator. The time evolution described by eq.~\eqref{eq:evolve} accounts for the polaron buildup. Let us consider first a four-wave mixing experiment \cite{borri:01}, where the quantum dot is initially brought into a superposition state $|0\rangle+|1\rangle$ and the polarization decay 
 $\left|\langle 1|U(t)\rho_0 e^{iH_0t}|0\rangle\right|$ through phonon dephasing is subsequently monitored. The gray lines in the inset of Fig.~2 show results of simulations with typical parameters for small quantum dots. The polarization amplitude drops on a picosecond timescale to a value that is several ten percent smaller than its initial value. At later times it remains constant, indicating the formation of a stable polaron. It was shown in Refs.~\cite{hohenester.jpb:07,hohenester.prb:06} that this initial drop is due to the emission of a phonon wavepacket away from the dot and the resulting imprint of the superposition state into the solid-state environment. The gray lines in the main panel of Fig.~2 report results of simulations where the system is initially brought into the superposition state $|1\rangle+|2\rangle$, corresponding to the buildup of a photon-exciton entanglement through the biexciton decay, see Eq.~\eqref{eq:entanglement}. The polarization amplitude $\left|\langle 2|U(t)\rho_0 U^\dagger(t)|1\rangle\right|$ slightly drops at early times due to the asymmetric phonon coupling through which the environment acquires information about the intermediate exciton state (see panels $X_H$ and $X_V$ in Fig.~1). However, because of the small $\eta$ value used in our simulations the resulting loss of entanglement is extremely small (less than a percent). 
 
\begin{figure}
\centerline{\includegraphics[width=0.8\columnwidth]{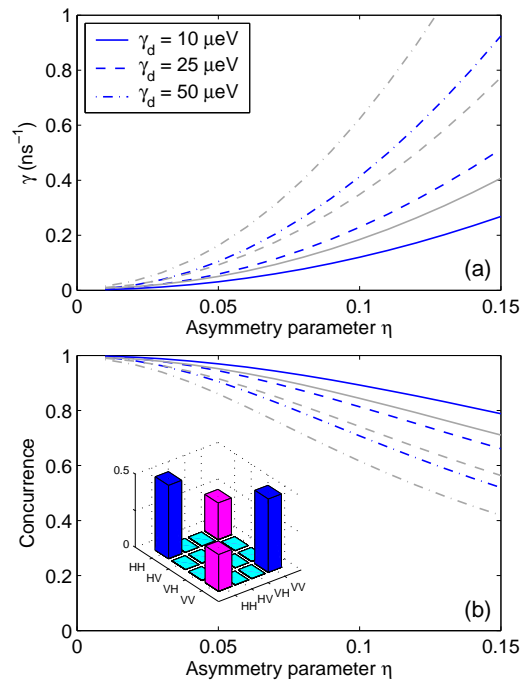}}
\caption{(color online) (a) Entanglement decay rate [see Eq.~\eqref{eq:two-photon}] as a function of asymmetry parameter $\eta$ for different elastic phonon scattering rates $\gamma_d$. The lattice temperature is 10 K (dark lines) and 20 K (gray lines). (b) Concurrence as a function of $\eta$ and for different values of $\gamma_d$ and lattice temperature, and for a radiative decay rate $\Gamma=1$ ns$^{-1}$. The inset reports a typical two-photon density matrix of the biexciton cascade, where the off-diagonal elements decay according to $\frac 12 e^{-\gamma\tau}$. The overall concurrence $1/(1+\gamma/\Gamma)$ results from this density matrix through time averaging over the whole decay process. 
}
\end{figure}

Things change considerably if additional phonon couplings are considered. Indeed, it is well known that the polarization amplitude further drops at later times \cite{borri:01}. This has been attributed to higher-order phonon processes \cite{muljarov:04}, extrinsic charging effects \cite{favero:07}, or elastic phonon scatterings at the device boundaries \cite{rudin:06}. The latter are particularly important for single quantum-dot devices, as the one considered in this work. Typical phonon linewidths due to elastic boundary scatterings are of the order of a few tens of $\mu$eV \cite{rudin:06}, where in this work we shall discard all directional and wavevector dependencies and use for simplicity a constant value $\gamma_d$. We model the elastic phonon scatterings within a master equation approach \cite{milburn:91}
\begin{equation}\label{eq:master}
  \dot\rho=-i[H,\rho]-\gamma_d\sum_\lambda\,
  [N_\lambda,[N_\lambda,\rho]]\,,
\end{equation}
with $N_\lambda=a_\lambda^\dagger a_\lambda^{\phantom\dagger}$. In accordance with elastic scatterings, this form conserves the number of phonon excitations and gives in thermal equilibrium an exponentially damped correlation function $\langle a_\lambda^\dagger(t)a_\lambda^{\phantom\dagger}(0)\rangle =e^{(i\omega_\lambda-\gamma_d)t}\bar n_\lambda$, where $\bar n_\lambda$ is the Bose-Einstein distribution. We solve the master equation \eqref{eq:master} numerically by discretizing the time domain into subintervals of length $\delta t$ and expanding $\rho$ and $U(\delta t)$ in the basis of phonon number states. The dark lines in Fig.~2 report results of simulations including elastic phonon scatterings, showing a constant decay of the polarization amplitude at later time. This is attributed to the combined effect of polaron formation and elastic phonon scatterings. In the formation of a polaron, the exciton becomes partially entangled with the phonon degrees of freedom. In turn, the elastic boundary scatterings of the phonons contributing to the polaron lead to an effective ``measurement'' of the environment on the superposition properties of the intermediate exciton state. Thereby the system becomes einselected from a maximally entangled to a mixed state. Thus, after a given time $\tau$, when the second photon is emitted in the exciton decay, the reduced degree of photon-exciton entanglement is swapped to a reduced photon-photon entanglement.

The decrease of the polarization amplitude with time turns out to be reasonably well described by an exponential decay with rate $\gamma$ [see also Eq.~\eqref{eq:two-photon}], which is reported in Fig.~3(a) for different asymmetry parameters and phonon broadenings $\gamma_d$. One observes that even small asymmetries $\eta$ in the phonon coupling result in a significant entanglement loss. For instance, using $\gamma(\eta=0.1)\sim 0.25$ ns$^{-1}$ at 10 K for a moderate phonon broadening $\gamma_d=25\,\mu$eV, we obtain for a typical radiative decay rate $\Gamma\sim 1$ ns$^{-1}$ a decrease of the concurrence by about twenty percent [see Fig.~3(b)]. Increased values of $\eta$, $\gamma_d$, or temperature even result in higher losses.


The mechanism described above might explain the small degree of photon entanglement observed in recent experiments \cite{stevenson:06}, and is at the same time compatible with the long measured spin scattering \cite{kroutvar:04} and dephasing \cite{greilich:06} times observed in large quantum dot samples. Theoretically, the magnitude of the phonon asymmetry might be estimated from calculations for realistic quantum dots \cite{seguin:05,bester:05}, including strain and piezoelectric fields. Experimentally, a time-resolved measurement of $G^{(2)}(\tau)$ would allow a direct observation of concurrence decay. Quite generally, the degree of photon entanglement can be enhanced by decreasing the radiative decay times through stronger microcavity couplings \cite{stace:03,troiani:06b}. In addition there might be other coupling channels to the environment, such as optical phonons or localized vibrations decaying through anharmonic processes, or impurity defects and charging centers in the vicinity of the dot \cite{favero:07} that produce fluctuating electric fields, which couple differently to the exciton states and thereby result in entanglement losses.


In summary, we have theoretically investigated the effects of environment couplings in the cascade decay of a biexciton confined in a single quantum dot. We have shown that a slightly polarization-dependent exciton-phonon coupling results, through the formation of a polaron in the vicinity of the dot, in a small loss of photon entanglement. This effect becomes drastically enhanced when the phonons contributing to the polaron scatter at the device boundaries. In consequence, the concurrence may easily drop by several ten percent even for moderate elastic phonon scatterings and low temperature. Our results suggest that in addition to the fine structure splitting, which can be controlled by external magnetic or electric fields, also extrinsic effects of the environment might provide a limiting factor in quantum-dot based entangled-photon sources.

We gratefully acknowledge helpful discussions with Filippo Troiani. This work has been supported in part by the Austrian Science Fund FWF under projet P18136--N13.

\end{document}